\documentstyle[12pt]{article}
\def\be{\begin{equation}}
\def\ee{\end{equation}}
\def\gsim{\mathrel{\raise.3ex\hbox{$>$\kern-.75em\lower1ex\hbox{$\sim$}}}}
\def\lsim{\mathrel{\raise.3ex\hbox{$<$\kern-.75em\lower1ex\hbox{$\sim$}}}}
\gdef\journal#1,#2,#3,#4.{{\it #1~}{\bf #2} (#4) #3}
\def\prd{\journal Phys. Rev. D,}
\def\prl{\journal Phys. Rev. Lett.,}
\def\npb{\journal Nucl. Phys. B,}
\def\plb{\journal Phys. Lett. B,}
\def\npb{\journal Nucl. Phys. B,}
\def\apj{\journal Ap. J.,}
\begin{document}
\begin{center}
{\Large \bf Polarization of the microwave background\\
in inflationary cosmology\footnote{Work partially supported by
Fundaci\'on Antorchas and by the DGXII of the Commission of the European
Communities under contract No. C 11-0540-M(TT).}
}\\
\vspace{.2in}
Diego D. Harari\footnote{Also supported by CONICET.}
and Mat\'\i as Zaldarriaga\\
\vspace{.2in}
Departamento de F\'\i sica, Facultad de Ciencias Exactas y Naturales\\
Universidad de Buenos Aires, Ciudad Universitaria - Pab. 1\\
1428 Buenos Aires, Argentina\\
\end{center}
\vspace{.3in}
\centerline{\bf ABSTRACT}
\noindent  We evaluate the large scale polarization of the cosmic
microwave background induced, via Thomson scattering prior to
decoupling, by nearly scale-invariant spectra of scalar and tensor
metric perturbations, such as those predicted by most inflationary
models of the early Universe. We solve the radiative transfer equation
for the polarized photon distribution function analytically, for
wavelengths of order or larger than the horizon size at decoupling,
assuming no reionization.
The induced polarization is proportional to the redshift
induced by the perturbations at decoupling and to the duration of the
decoupling transition.
Normalizing the spectra to the quadrupole anisotropy measured by the
COBE satellite, the expected degree of linear polarization is  $P\lsim
10^{-7}$, more than two orders of magnitude smaller than current
bounds. The dependence of $P$ with the tilt in the spectrum away from
scale invariance is too weak to be of observational relevance or to
provide a sensitive discrimination between scalar and tensor
fluctuations. The variation of $P$ with the angular scale is
significant.
\begin{center}
\vspace{.3in}
{\it To appear in Physics Letters B}
\end{center}
\newpage

The most plausible explanation of the quadrupole anisotropy in the
temperature of the cosmic microwave background measured by the COBE
satellite \cite{COBE92}, is that it originates in redshifts induced by
metric fluctuations around a Robertson-Walker background, through the
Sachs-Wolfe effect \cite{Sachs67}.

The fluctuations in the metric may be scalar (energy-density) or
tensor perturbations (gravitational waves). Inflationary cosmological
models predict them both, as a result of quantum fluctuations at early
stages. Exponential inflation predicts a scale-invariant, gaussian
spectrum of scalar fluctuations \cite{scalar}, and a smaller amount of
tensor fluctuations \cite{tensor}. Other inflationary models, for
instance power-law inflation \cite{Abbott84}, predict spectra slightly
tilted away from scale invariance. It would be of great value, and
would shed much light upon the Universe early evolution, to know what
fraction of the quadrupole anisotropy measured by COBE is due to
scalar fluctuations and what fraction is due to gravitational waves.
Most inflationary models predict a ratio between scalar and tensor
fluctuations proportional to the tilt in the spectrum away from scale
invariance. In principle, independent measurements of the relative
amplitudes of scalar and tensor fluctuations and of the spectral
index, could thus provide significant evidence for
inflation \cite{Krauss92,Davis92,Mollerach92,Salopek92,Liddle92}, if
they happen to fit the predicted ratio.  Put it another way,
independent measurements of these parameters would allow a
determination of the potential of the field driving
inflation \cite{Copeland93}. This program  however, is not easy to carry
out in practice, among other things because of the statistical nature
of the predictions (the ``cosmic variance'' problem), and because
discrimination between tensor and scalar sources of anisotropy
requires measurements at small angular
scales \cite{Davis92,Crittenden93a}, where the primordial spectrum is
affected by many different processes that strongly depend on the
details of the cosmological model.

Its degree of polarization is another (as yet unmeasured) property of
the microwave background radiation, which determination would shed
more light upon the Universe early evolution. The polarization of the
microwave background radiation is actually tightly related to its
anisotropy: Thomson scattering of anisotropic radiation by free
electrons before recombination induces some degree of linear
polarization \cite{Rees,Basko}.

In this paper we evaluate the degree of linear polarization induced both by
scalar as well as tensor fluctuations, with a gaussian, power law spectrum,
not far away from the scale invariant (Harrison-Zeldovich) type.  Our analysis
is based on the analytic method and approximations developed by Basko and
Polnarev  to study the polarization induced by anisotropic expansion of the
Universe \cite{Basko} and by gravitational waves \cite{Polnarev}. We extend
their method to include the effect of scalar fluctuations. Closely following
their approach to the subject, we solve the radiation transfer equations for
the polarized photon distribution function analytically, performing
approximations which will prove valid for the large angular scale ($\gsim
1^{\rm o}$) polarization. We assume that the Universe is already
matter-dominated at recombination, and that there is no further reionization.
Polarization measurements of the cosmic microwave background at large angular
scales could very well serve as a test of the reionization
history \cite{Basko,Ng93,Crittenden93b}, but this is not the focus of the
present paper. Instead, we concentrate (assuming no reionization) on the
predictions of inflationary models, such as power law inflation. In other
words, we normalize the scalar and tensor fluctuations spectra to the COBE
measurements of the anisotropy in the microwave background temperature, and
determine the expected large angular scale polarization. From Polnarev's
results \cite{Polnarev} one expects a rough estimate on the degree of
polarization of the microwave background of around a few percent of its large
scale anisotropy, and this is a rather small number, about two orders of
magnitude smaller than current bounds. We feel, however, that the recent
measurement of the quadrupole anisotropy justifies a more detailed analysis of
the inflationary predictions for the large angular scale polarization.
Our motivation is to find the dependence of the large scale
polarization with the spectral index of the fluctuations, which if large could
in principle provide an adittional indirect test for inflation. Other
dependences of the degree of polarization, such as upon the angular scale or
the cosmological parameters, are also of interest.

Our work partially overlaps with a recent preprint by Crittenden, Davis and
Steinhardt \cite{Crittenden93b}, where the same subject was treated
numerically. Our approach, valid only for large angular scales under the
assumption of no reionization, is of a more limited applicability than their
full numerical treatment. On the other hand, our analytic results, based on
the application of Basko and Polnarev's method \cite{Basko,Polnarev}, provide
some interesting insights into the problem, such as the explicit dependence
upon the spectral index, the angular scale and upon  parameters of the
cosmological model.

The polarization of the cosmic
microwave background radiation will be determined from the Stokes parameters
that result from solving the Boltzmann equation in a perturbed spatially-flat
Robertson-Walker background. The metric reads
\be ds^2=a^2(\tau)[-
d\tau^2+(\delta_{ij}+h_{ij})dx^idx^j]\quad , \label{metric}
\ee
where $\tau$ is
the conformal time, related to proper time $t$ by $d\tau=dt/a(t)$,
$|h_{ij}|<<1$, and $i,j=1,2,3$. The polarized photon distribution function is
represented, in a general case, by a column vector ${\bf\hat
n}(\nu,\theta,\phi)= (I_l,I_r,U,V)$, its components being the Stokes
parameters of the radiation that arrives from a direction defined by the
angles $\theta,\phi$ with frequency $\nu$ \cite{Chandrasekar}. $I_l,I_r$ are
the intensities in two  orthogonal directions defined by the unit vectors
$\hat\theta$ and $\hat\phi$ of spherical coordinates. The Boltzmann equation
reads
\begin{equation}
\left({\partial \over \partial \tau}+e^i {\partial
\over \partial x^i}\right) {\bf\hat n} =-{d\nu \over d\tau}{\partial {\bf\hat
n} \over \partial\nu}- \sigma_T N_e a\left[{\bf\hat n}-\frac{1}{4\pi} \int_{-
1}^1 \int_0^{2\pi}
 P(\mu,\phi,\mu^\prime,\phi^\prime){\bf\hat n}d\mu^{\prime}
d\phi^{\prime}\right]\label{eqbol}
\end{equation}
Here $P(\mu,\phi,\mu^\prime,\phi^\prime)$ is a $4\times 4$ matrix that
characterizes the
effect of Thomson scattering on polarization \cite{Chandrasekar},
$\mu=\arccos\theta$, $e^i$ is a unit vector in the direction of
$(\theta,\phi)$,
$\sigma_T$ is the Thomson scattering cross section and $N_e$ is the free
electron density. The effect of the metric perturbations on the frequency of
the photons is given by the Sachs-Wolfe formula, which in the
synchronous gauge reads \cite{Sachs67}
\begin{equation}
\frac{1}{\nu}\frac{d\nu}{d\tau}=\frac{1}{2}{\partial h_{ij}\over \partial
\tau}e^ie^j\label{SW}
\end{equation}
The Stokes parameter $V$ actually decouples from the rest, and we will not
take it into further consideration, since it is not necessary to
determine the degree of linear polarization, given by $P=\sqrt{Q^2 +
U^2}/I$,   with $I=I_l+I_r$, $Q=I_l-I_r$.

We describe the growing mode of adiabatic scalar fluctuations
in a matter-dominated,
spatially-flat Robertson-Walker background, in terms of
Bardeen's gauge-invariant variable $\zeta$, as \cite{Bardeen}
\begin{equation}
h_{ij}(\vec x,\tau)=-{1\over 15}\tau^2{\partial^2\zeta(\vec x)\over\partial
x^i\partial x^j}+{1 \over 5} \zeta(\vec x) \delta_{ij}\quad .
\end{equation}
$\zeta$ coincides with the energy-density fluctuations $\delta\rho/\rho$
in a matter-dominated regime for wavelengths inside the horizon. We decompose
it in plane waves, $\zeta(\vec x)=\int d^3k e^{i\vec k\cdot\vec x}\zeta(\vec
k)$, where $\zeta(\vec k)$ is a gaussian, random variable with
statistical average $\langle\zeta(\vec k)\zeta(\vec
q)\rangle=P_\zeta(k)\delta(\vec k - \vec q)/4\pi k^3$. $P_\zeta$
defines the
spectrum of scalar fluctuations, which we will take to be of the form
\be
P_\zeta (k)=P_\zeta k^{n-1}
\label{Pzeta}\ee
where $P_\zeta$ is a constant, and the spectrum is scale invariant
(constant amplitude on scales equal to the horizon at any given
time) if $n=1$.

Analogously, tensor fluctuations in the transverse traceless gauge,
that enter the horizon during a matter-dominated regime, are described by
\be
h_{ij}(\vec x,\tau)=\sum_\lambda\int d^3 ke^{i\vec k\cdot\vec x}
h_\lambda(\vec k)\epsilon_{ij}(\lambda,\vec k)\ \left( {3j_1(k\tau)\over
k\tau}\right)
\ee
where $\lambda=1,2$ denotes the gravitational wave
polarization, characterized by $\epsilon_{ij}$,
$j_1$ are spherical Bessel functions, and $h_\lambda$ are random
variables with statistical average
$\langle h_\lambda(\vec k,\tau)h_{\lambda^\prime}(\vec
q,\tau)\rangle=P_h(k)\delta(\vec k - \vec q)\delta_{\lambda\lambda^\prime}
/4\pi k^3$. We will consider a power spectrum  of the form
\be
P_h (k)=P_h k^{n-1}\quad ,
\label{Ph}\ee
which is scale invariant if $n=1$.

We first discuss the polarization induced by one single scalar mode, of the
form $\zeta(\vec x)=\zeta(k)e^{i\vec k\cdot\vec x}$. Since
the problem has rotational symmetry around the direction of $\vec k$, only two
Stokes parameters need to be considered $(U=0)$. Now ${\bf\hat n}$ is a
two-component
column vector, which we expand to first order in the perturbation, as
${\bf\hat n}={\bf\hat n_0}+(1/2)n_0{\bf\hat n_1}(\vec x,\tau,\nu)$,
with ${\bf\hat n_0}=(1/2)n_0(1\ 1)$, where $n_0(\nu)$ is the blackbody
distribution. Boltzmann's equation
(\ref{eqbol})  simplifies considerably:
\begin{equation}
\left({\partial \over \partial \tau}
+e^i {\partial \over \partial x^i}\right)
{\bf\hat n}_1 =-p F \mu^2
\pmatrix{1\cr 1\cr}
-q\left[{\bf\hat n}_1-\frac{3}{8} \int_{-1}^1
\pmatrix{2(1-\mu^2)(1-\mu^{\prime 2})+\mu^2 \mu^{\prime 2}&\mu^2\cr
\mu^{\prime 2}&1\cr}
{\bf\hat n}_1 d\mu^{\prime} \right]
\label{bolhom}
\end{equation}
where $q=\sigma_T N_e a$,
$p={d\ln n_0 \over d\ln \nu}\approx 1$ in the Rayleigh-Jeans zone
of the blackbody spectrum, and $F_S\equiv -\tau\zeta(k)
e^{i\vec k\cdot \vec x}k^2/15$ arises from the redshift induced by the
perturbation, as given by the Sachs-Wolfe equation (\ref{SW}).
We will solve  equation (\ref{bolhom})
in the long wavelength limit, neglecting the
spatial derivative in the l.h.s., but
anyhow keeping the $\vec k\cdot\vec x$-dependence in the distribution function.
This amounts to a quasi-stationary approximation \cite{Polnarev}, valid for
perturbations with wavelength much longer than the photon mean-free path
just prior to decoupling $(k<<q)$,
and also much longer than the distance that a photon
can travel during the decoupling transition, as we will
discuss in more detail later.
We normalize the present conformal time $\tau_0=1$,
then $a(\tau)={2\tau^2 \over H_0}$, and
$q\approx 0.14 \Omega_b h X_e (1+z)^2$, with $\Omega_b$ the density in
baryons in units of the critical energy density, $X_e$ the ionization
fraction, and
$H_0=100\ h\ \rm km\ seg^{-1}\ Mpc^{-1}$. Thus around decoupling,
$q\approx 10^4$. Notice that with our normalization $k\approx 2\pi$
corresponds to a wavelength comparable to the present horizon, and
$k\approx 30(2\pi)$ to the horizon at decoupling.

We solve eq. (\ref{bolhom}) through the
following ansatz for ${\bf \hat n_1}$:
\begin{equation}
{\bf\hat n}_1= \alpha(\tau){\bf\hat a}+\beta(\tau){\bf\hat b}+\gamma(\tau)
{\bf\hat c}
\label{abc}
\end{equation}
where
\begin{equation}
{\bf\hat a}=(\mu^2-{1\over 3})\pmatrix{1\cr 1\cr} \quad ;\quad
{\bf\hat b}=(1-\mu^2) \pmatrix{1\cr -1\cr} \quad ;\quad
{\bf\hat c}= \pmatrix{1\cr 1\cr}
\label{a1}
\end{equation}
With this ansatz, and neglecting the spatial derivatives in the l.h.s. of the
Boltzmann equation (\ref{bolhom}), we derive the following equations
\begin{equation}
{\partial\xi \over \partial \tau}+q\xi=F_S\quad ;\quad
{\partial\beta \over \partial \tau}+{3 \over 10}q
\beta=-{q \over 10}\xi\quad ;\quad
{\partial\gamma \over \partial \tau}={1\over 3}F_S
\label{ab}
\end{equation}
where we have defined the variable $\xi=\alpha+\beta$,
which decouples from $\beta$, and is a measure of the
anisotropy in the photon distribution function. $\beta$ instead is a
measure of the polarization, since $I_l-I_r=n_0\beta(1-\mu^2)$, and
the anisotropy, as measured by $\xi$, is its source. $\gamma$, on
the other hand, decouples from the rest, and just reflects a local adjustment
of the black body intensity, independent of the direction of propagation.
The equations for $\xi$ and $\beta$ can be formally integrated as
follows \cite{Polnarev}
\begin{equation}
\xi (\tau)=\int_0^{\tau}F_S(\tau^{\prime})e^{-\kappa(\tau,\tau^{\prime})}
d\tau^{\prime}
\end{equation}
\begin{equation}
\beta (\tau)=-{1\over 10}\int_0^{\tau}q(\tau^{\prime})\xi(\tau^{\prime})e^{-
{3\over 10}\kappa(\tau,\tau^{\prime})}
d\tau^{\prime}
\label{eqbeta2}
\end{equation}
where $\kappa(\tau,\tau^{\prime})=
\int_{\tau^{\prime}}^{\tau}q(\tau^{\prime \prime})d\tau^{\prime \prime}$
is the optical depth.
We wish to evaluate $\beta (\tau_0)$ from eq.(\ref{eqbeta2}).
This equation shows that the present polarization of the microwave background
(assuming no reionization) is essentially that produced at the times
when $qe^{-{3\over 10}\kappa(\tau_0,\tau^\prime)}$ is significantly different
from zero, {\it i.e.} around the time of recombination,
since much later the free electron density (and thus $q$)
is negligible, while much earlier the optical depth is very large.
To be more precise, the present polarization is the result of Thomson
scattering around the time of decoupling of matter and radiation, which occurs
slightly after the free electron density starts to drop significantly.
Although these values depend on the particular cosmological model considered
(through $\Omega,\ \Omega_b,\ H_0$, etc.), typically recombination occurs
around $z\approx 1400$ while decoupling  takes place between
$z\approx 1200$ and $z\approx 900$. This is the interval during which the
differential visibility differs significantly from zero, and is actually well
approximated by a gaussian \cite{Bond88}. We will denote by $\tau_d$
the conformal time around which decoupling occurs, and $\Delta\tau_d$ the
duration of the process. These quantities actually do not depend very strongly
on the cosmological parameters, typical values being (with the convention
$\tau_0=1$), $\tau_d\approx
3\times 10^{-2}$ and $\Delta\tau_d\approx 3\times
10^{-3}$ \cite{Bond88,Sunyaev}.

As explained above, we only need to know $\xi(\tau)$ around $\tau_d$
in order to integrate eq.(\ref{eqbeta2}). If we define
$\kappa(\tau)\equiv \kappa(\tau_0,\tau)$, then
$\kappa(\tau,\tau^{\prime})=\kappa(\tau^{\prime})-\kappa(\tau)$, and
we may approximate around $\tau_d$:
\begin{equation}
\xi(\tau)\approx F_S(\tau_d)e^{\kappa(\tau)}\int_0^{\tau}
e^{-\kappa(\tau^{\prime})} d\tau^{\prime}
\approx F_S(\tau_d)\Delta\tau_d e^{\kappa(\tau)} E(\kappa(\tau))
\label{E}
\end{equation}
where in the last step we made the approximation
${d\kappa \over d
\tau^{\prime}} \approx {-\kappa(\tau^\prime) \over \Delta \tau_d}$,
which is justified by the almost gaussian shape of the differential visibility
around decoupling, and we defined
$E(\kappa)=\int_1^\infty {e^{-\kappa x} \over x} dx$.
Then, using that $-q d\tau = d\kappa$, we integrate $\beta$:
\begin{equation}
\beta \approx -{1\over 10}F_S(\tau_d) \Delta\tau_d \int_0^{\infty}e^{-
{3\over 10}\kappa} e^{\kappa} E(\kappa)
d\kappa \approx - 0.17 F_S(\tau_d) \Delta\tau_d\quad .
\end{equation}
Finally, the polarization produced by a single scalar mode is given by
\begin{equation}
P^2_S=\vert\beta\vert^2 (1-\mu^2)^2\quad .
\end{equation}

The polarization induced by a single tensor mode of a given
polarization can be calculated in a
similar way. Now, as opposed to the scalar case, there is no longer
rotational invariance around the direction of $\vec k$, thus $U\ne 0$,
and  ${\bf\hat n}$ is a three column vector
$(I_l,\ I_r,\ U)$, which we expand  as
${\bf\hat n}={\bf\hat n_0}+(1/2)n_0{\bf\hat n_1}(\vec x,\tau,\nu,\phi)$,
with ${\bf\hat n_0}=(1/2)n_0(1\ 1\ 0)$.
The angular dependence of the redshift produced by the gravitational wave
due to the Sachs-Wolfe effect suggests an ansatz for
${\bf\hat n_1}$ of the form: \cite{Polnarev}
\begin{equation}
{\bf\hat n}_1= \alpha(\tau){\bf\hat a}+\beta(\tau){\bf\hat b}
\end{equation}
where
\begin{equation}
{\bf\hat a}=(1- \mu^2) \cos(2\phi)\pmatrix{1\cr 1\cr 0\cr} \quad ;\quad
{\bf\hat b}=\pmatrix{(1+\mu^2)\cos(2\phi)\cr -(1+\mu^2)\cos(2\phi)\cr
4\mu\sin(2\phi)\cr}
\label{ab1}
\end{equation}
This corresponds to a gravitational wave with polarization $h_+$.
The formulae for a polarization $h_\times$ are identical to these after
interchange of $\cos(2\phi)$ with $\sin(2\phi)$.
With this ansatz, the equations for $\zeta=\alpha+\beta$ and $\beta$
are exactly the same as those in eq.(\ref{ab}) for the
scalar case, but with $F_S$ replaced by
$F_T\equiv {1\over 2}  e^{i\vec k\cdot\vec x}
h_\lambda(\vec k)\ {d\over d\tau}{3j_1(k\tau)\over k\tau}$.
Under the same approximations assumed valid for the scalar case,
the result for $\beta$ is then
$\beta\approx 0.17 F_T(\tau_d) \Delta\tau_d$. Finally, the polarization
induced  by  a tensor mode is given by
\begin{equation}
P^2_T=\vert\beta\vert^2 [(1+\mu^2)^2+4\mu^2]
\end{equation}
where we have already summed over the two polarizations of the gravitational
wave.

Both for scalar as well as tensor fluctuations, $\beta\propto
F(\tau_d)\Delta\tau_d$, with the appropriate $F$ in each case.
This result for $\beta$, which is a measure of the induced
polarization, has an intuitive explanation, after noticing that $F$
measures the anisotropy in the rate of change in the frequency of
the cosmic background photons. Indeed, the Sachs-Wolfe formula for the
scalar and tensor modes can be written as
\begin{equation}
{1\over \nu} {d\nu\over d\tau}=F_S\ \mu^2 \quad;\quad {1\over
\nu} {d\nu\over d\tau}=F_T\ (1-\mu^2) \cos(2\phi) \quad .
\end{equation}
This anisotropy in the redshift translates into a difference in the number of
photons of a given frequency traveling in different directions,
and this is what makes possible a net polarization after Thomson scattering.
Besides, the polarization builds up over a period $\Delta\tau_d$ only, since
anisotropies in the photon distribution produced much before decoupling are
erased by successive  scatterings, while after decoupling scattering is
negligible and thus polarization can no longer be induced.

Next we evaluate the rms total polarization that results from the
superposition of all the statistically independent contributions
of scalar and tensor modes respectively:
\begin{equation}
\begin{array}{ll}
& \langle P^2\rangle_S=\int d^3k\ (1-\mu^2)^2 \vert\beta\vert^2
={8\over 15}\int_0^{k_{max}} {dk\over k}\ (P_{\zeta}
k^{n-1})[0.17
\Delta\tau_d {k^2\tau_d\over 15}]^2\\
& \langle P^2\rangle_T=\int d^3k\ [(1+\mu^2)^2+4\mu^2]\vert\beta\vert^2
={4\over 5}\int_0^{k_{max}} {dk\over k}\ (P_h k^{n-1})[0.17
\Delta\tau_d\ {d\over d\tau}({3j_1(k\tau)\over k\tau})\vert_{\tau_d}]^2
\end{array}
\end{equation}
We have introduced a cut-off, $k_{max}$, determined by the minimum
wavelength to which a given experimental setup is sensitive.
In general, measurements result from an
average over a certain angular scale, for instance the aperture of
the antenna's horn. With our conventions, a wavelength with comoving
wavevector $k$ subtends an angle $\theta$ such that
$k\approx \pi/\sin(\theta/2)$. This relation determines $k_{max}$
for measurements averaged over an angle $\theta$.

Finally, for scalar perturbations
\begin{equation}
\langle P^2\rangle_S={8\over 15}\ {P_{\zeta}k_{max}^{n-1}\over (n+3)}[0.17
\Delta\tau_d {k_{max}^2\tau_d\over 15}]^2
\end{equation}
while the result for tensor perturbations can be written as
\be
\langle
P^2\rangle_T= D_p(n,k_{max})\langle P^2\rangle_S\quad ,
\ee
where
\begin{equation}
D_p\equiv 13.5\ g(n,k_{max}) {P_h\over P_{\zeta}}\quad;\quad g(n,k_{max})\equiv
25
(n+3){\int_0^{(k_{max}\tau_d)}dx\ x^n\ \left({d\over dx}\left(3j_1(x)\over
x\right)\right)^2\over (k_{max}\tau_d)^{n+3}}\ .
\end{equation}
With this definition, $g(n,k_{max})\to 1$ when $k_{max}\tau_d\ll 1$, {\it
i.e.} for wavelengths well outside the horizon at decoupling. The total rms
polarization is given by the sum of the contributions due to scalar and tensor
modes, given their statistical independence:
\begin{equation}
\sqrt{\langle P^2\rangle}=\sqrt{\langle P^2\rangle_S+\langle P^2\rangle_T}
\end{equation}

Now we assume that the scalar and tensor perturbations have as common origin a
period of power law inflation, in which case the scalar and tensor
power spectra are related
by $P_h/P_{\zeta}\approx 4(n-1)/9 (n-3)$ \cite{Davis92,Mollerach92,Salopek92}.
The ratio of the tensor and scalar contribution to the
quadrupole moment in the temperature anisotropy
can also be approximated by a simple
function of $n$, $\langle a^2_2\rangle_T/\langle a^2_2\rangle_S \equiv
D_a\approx 13(n-1)/(n-3)$ \cite{Mollerach92}. Assuming that the anisotropy
measured by COBE is due to these inflation-produced perturbations alone,
then it is simple to
express both   $P_{\zeta}$ and $P_h$ in terms of the measured quadrupole
and the spectral index.  For instance:
\begin{equation}
P_{\zeta}={45\over 4\pi^2}\ {2^{1-n} (3-n)\over(8-7n)}\ {\Gamma((9-n)/2)\
\Gamma^2(2-n/2) \over \Gamma((3+n)/2)\ \Gamma(3-n)}\ \langle a^2_2\rangle
\end{equation}
with $\langle a^2_2\rangle=\langle a^2_2\rangle_S+\langle a^2_2\rangle_T
\approx 2\times 10^{-5}$, the quadrupole measured by COBE.
Finally, the total rms polarization predicted by power-law inflationary
models, as a function of the spectral index $n$, is given by
\begin{equation}
P\approx 1.2\times 10^{-7}\ \left({\Delta\tau_d \over 3\times 10^{-3}}\right)\
\left({k_{max}\over 50}\right)^2\ \left({\tau_d} \over3 \times 10^{-2}
\right)  \left({\sqrt{\langle a^2_2\rangle}\over 2\times 10^{-5}}\right)\
C(n,k_{max})
\end{equation}
with
\begin{equation}
C^2\equiv{8\over 3(n+3)\pi}k_{max}^{n-1} {\Gamma((9-
n)/2)\ \Gamma^2(2-n/2) \over 2^{n-1}\Gamma((3+n)/2)\ \Gamma(3-n)}
\left(1+D_p\over
1+D_a\right)
\end{equation}
The function $C$ has been defined such that $C=1$ if $n=1$.
The value  $k_{max}\approx 50$ corresponds to measurements that average
over an angular scale of order $\theta\approx 7^o$.

In Fig. 1 we plot the total rms polarization for $k_{max}=50$,
as well as the individual tensor
and scalar contributions, for values of $n$ between $0.5$ and $1$,
those consistent with COBE measurements. The
predicted degree of polarization is more than two orders of magnitude smaller
than current bounds, $P< 6\times 10^{-5}$ \cite{Lubin}, obtained with a $7^o$
aperture horn at $33$ GHz.
The polarization induced by
tensor modes vanishes if $n=1$ and equals the scalar contribution when
$n=0.5$.
The total polarization  decreases by a factor about 0.6 as $n$ varies from 1 to
0.8 ($n\approx 0.8$ corresponds to equal contributions of scalar and tensor
modes to the large scale anisotropy), and about 0.4 between $n=1$ and $n=0.5$.
Notice that the dependence of the total polarization upon $n$ is mainly due to
the tilt in the spectra, {\it i.e.} it comes mostly from the term
$k_{max}^{(n-1)/2}$. This is because polarization is proportional to the
redshift induced by the perturbations at decoupling, and is no longer produced
afterwards, while anisotropy continues to build up.  Consequently, the
dominant wavelengths for polarization are much smaller than those that
dominate the large scale anisotropy. Once the spectra are normalized to the
large scale anisotropy measured by COBE, smaller $n$ implies less power on
smaller scales.

A large class of inflationary models predict a definite ratio
between the spectral index $n$ and the scalar and tensor contributions to the
anisotropy in the microwave background \cite{Davis92,Mollerach92}. We have
shown, applying the methods developed by Basko and
Polnarev \cite{Basko,Polnarev} to the specific spectra predicted by
inflationary models, that there is also a definite relation between
the net polarization and $n$, at least assuming no reionization.
Unfortunately, the rather weak dependence of $P$ with $n$ makes it unlikely
that measurements of polarization could provide a sensitive confirmation of
these predictions. Our conclusions in this regard are similar to those of
Crittenden, Davis and Steinhardt \cite{Crittenden93b}, that in a recent
preprint approached the same subject numerically. Our analytic results,
on the other hand,  are only valid if
$k_{max}\Delta\tau_d<<1$, which means $k_{max}<<300$, while the numerical
treatment of Ref. \cite{Crittenden93b} applies to a more general situation.

The large scale polarization depends upon cosmological parameters, such as
$\Omega, \Omega_b, H_o$, etc., only through $\tau_d$ and $\Delta\tau_d$,
{\it i.e.} through the parameters that characterize the decoupling transition,
and this is a very weak dependence \cite{Bond88}.

Notice finally that $P$ is proportional to $k_{max}^2$.
This is a consequence of the fact that the polarization induced by a single
mode is proportional to the redshift it causes at decoupling, irrespective of
its wavelength, and because in the linear regime the perturbations grow as
$(k\tau)^2$. Thus, the dependence of the degree of polarization with
the angular scale over which polarization data is averaged is significant.



\noindent{\bf Figure Caption}

\noindent Fig. 1: Total rms large scale ($\approx 7^o$)
polarization of the cosmic microwave background
$P=\sqrt{P_T^2+P_S^2}$ and individual scalar and tensor contributions,
$P_S$ and $P_T$, as a function of the spectral index $n$ in power-law
inflationary models, assuming no reionization.

\end{document}